\documentclass[12pt,a4paper]{article}
\usepackage{axodraw}
\newcommand{\be}{\begin{equation}}
\newcommand{\ee}{\end{equation}}
\newcommand{\ba}{\begin{eqnarray}}
\newcommand{\ea}{\end{eqnarray}}

\newcommand{\bea}{\begin{eqnarray*}}
\newcommand{\eea}{\end{eqnarray*}}
\newcommand{\bet}{\begin{center} \begin{tabular}}
\newcommand{\ent}{\end{tabular} \end{center}}
\newcommand{\bary}{\begin{array}}
\newcommand{\eary}{\end{array}}
\newcommand{\bit}{\begin{itemize}}
\newcommand{\eit}{\end{itemize}}

\newcommand{\veps}{\varepsilon}
\newcommand{\sla}[1]{#1 \!\!\!/}

\newcommand{\non}{\nonumber}

\newcommand{\dm}{\delta m}
\newcommand{\MSb}{$\overline{\mathrm{MS}}$ }

\newcommand{\adp}{\frac{\alpha_s}{4 \pi} }

\newcommand{\Fh}[2]{\,{}_#1F_#2}
\newcommand{\Fs}[3]{\!\!\left[\begin{array}{c}#1\,;\\#2\,;\end{array}#3\right]}

\newcommand{\Fup}[2]{\Fs{#1}{#2}{\frac{q^2}{m^2}}}

\newcommand{\Li}[1]{\,{\rm Li}_{#1}}

\newcommand{\bb}{\begin{thebibliography}}
\newcommand{\eb}{\end{thebibliography}}

\newcommand{\lab}[1]{\label{#1}}
\newcommand{\re}[1]{(\ref{#1})}
\newcommand{\lz}{\ln\bar z}
\newcommand{\bz}{\bar z}
\newcommand{\z}{&\hspace*{-8pt}}

\begin{document}
%=========== title page ======================
\thispagestyle{empty}
\onecolumn
%\date{\today}
\vspace{-1.4cm}
\begin{flushleft}
%{DESY 98-026 \\}
%{hep-ph/980????\\}
%March, 1998

\end{flushleft}
\vspace{1.5cm}

\begin{center}

{\LARGE {\bf
Two-loop QCD corrections of the massive fermion propagator $^{1)}$}\vglue 10mm
        }

\vspace{1.5cm}

\vfill
{\large
 J. Fleischer$^{a}$, F. Jegerlehner$^{b}$,
 O.V. Tarasov$^{a,b,}$\footnote{On leave of absence from JINR,
141980 Dubna (Moscow Region), Russian Federation.}$^,\!\!$
\footnote{
Supported by BMBF
under contract PH/05-7BI92P 9
}
  and O.L. Veretin$^{a,}$%
\addtocounter{footnote}{-0}%
\footnotemark[\value{footnote}]%
}

\vspace{2cm}
$^a$~~Fakult\"at f\"ur Physik~~~~~\\
Universit\"at Bielefeld \\
D-33615 Bielefeld, Germany

\vspace{1cm}

$^b$~~Deutsches Elektronen-Synchrotron DESY~~~~~ \\
Platanenallee 6, D--15738 Zeuthen, Germany\\
\end{center}

\vfill

\begin{abstract}
The off-shell two-loop correction to the massive quark propagator in
an arbitrary covariant gauge is calculated and results for the bare
and renormalized propagator are presented. The calculations were
performed by means of a set of new generalized recurrence relations
proposed recently by one of the authors~\cite{OVT97}. From the position
of the pole of the renormalized propagator we obtain the relationship
between the pole mass and the \MSb mass. This relation
confirms the known result by Gray et al.~\cite{GBGS}. The bare
amplitudes are given for an arbitrary gauge group and for arbitrary
space-time dimensions.

\end{abstract}

%\vfill
\vfill\footnoterule\noindent
$^1$) {\sl In memoriam} Leonid Viktorovich Avdeev, 1959-1998\\
\newpage

\setcounter{footnote}{0}

\section{Introduction}
In electroweak precision physics there is ongoing interest to improve
the status of precise theoretical predictions of the Standard
Model. Attempts are being made by several groups to provide more
complete two-loop calculations. The major obstacles in doing so are
still lacking methods to perform accurate numerical calculations of
two-loop integrals. Unlike in the one-loop case integrals cannot be
performed analytically in general. Recently substantial progress was
made in the development of efficient algorithms which allow us to
calculate two--loop diagrams for arbitrary masses in a canonical
way~\cite{OVT97}.  In the present paper we describe a simple
application of these techniques: the calculation of the two-loop
quark propagator in Quantum Chromodynamics (QCD) with one massive quark
and the others massless. Bare expressions are
obtained for an arbitrary gauge group and for arbitrary space-time
dimensions. The knowledge of the full off-shell propagator is of
particular interest for the discussion of the scheme dependence of
radiative corrections, since it allows us to relate not only the pole
mass $M$ to the \MSb mass $m$ but also to the so called momentum subtraction
schemes~\cite{CelmasterGonsalves} (MOM) and the Euclidean mass
definition~\cite{GeorgiPolitzer}. In lattice QCD a MOM subtraction
scheme was recently proposed for the non-perturbative operator
renormalization with matching conditions to perturbation theory far
off-shell~\cite{Martinelli}. These considerations are of great
physical interest in particular in the context of decoupling of heavy
particles and the effective QCD schemes where it is required to relate
QCD in the \MSb scheme for effective theories with different number of
(light) flavors~\cite{Decoupling}.

The two-loop corrections to the electron propagator were first
considered in Quantum Electrodynamics (QED)~\cite{Sabry}. In QCD so
far only the relationship between the pole mass and the
\MSb mass was investigated to two-loops~\cite{GBGS}.
A similar on-shell calculation, within the dimensional reduction
scheme, was given in~\cite{AK}. A detailed discussion of the mass
renormalization was presented in~\cite{Coqu} for QED, in~\cite{FJ}
for the electroweak Standard Model and in~\cite{Bin} for grand unified
theories.

In sect.~2, we introduce some notation for the quark propagator. The
calculation and the results for the renormalized amplitudes are
discussed in sect.~3, while sect.~4 is devoted to the relationship
between the pole mass and the \MSb mass. Technical details and bare
amplitudes will be given in a few appendices.

\section{The quark propagator}
The renormalization of the fermion propagator is conveniently discussed by
looking at the full inverse bare propagator first
\be
S_{\rm F}^{-1}(q)=\sla{q} -m_0-\Sigma(q)\,,
\ee
where $\Sigma(p)$ is the one-particle irreducible fermion self-energy.
We write the inverse bare propagator as
\be
S_{\rm F}^{-1}(q)=\sla{q} B_{\rm bare} -m_0 A_{\rm bare}
\ee
in terms of two dimensionless scalar amplitudes $A_{\rm bare},B_{\rm
bare}$, which are functions of the four-momentum square $q^2$, the
bare mass $m_0$, the bare QCD coupling constant $\alpha_{s0}\equiv
g_0^2/(4\pi)$ and the bare gauge parameter $\xi_0$.  Off the
mass-shell, when $q^2 \neq M^2$, the bare amplitudes are {\em
infrared} {\em finite} but ultraviolet singular. As usual we
start from dimensionally regularized bare amplitudes. Renormalization
of the fermion propagator is performed order by order in perturbation
theory by writing the bare mass as a sum of the renormalized mass plus
a mass counterterm
\be
m_0=m+\dm=Z_m\:m\,,  \qquad  Z_m \equiv 1+\frac{\dm}{m}
\ee
and, correspondingly, for the bare strong coupling constant
\be
\alpha_{s0}=\alpha_s+\delta \alpha_s = Z_g\:\alpha_s\,,  \qquad
Z_g\equiv 1+\frac{\delta \alpha_s}{\alpha_s}\,,
\ee
while for the bare gauge parameter we have
\be
\xi_0= Z_3\xi\,.
\ee
We define $\xi$ such that $\xi=1$ in the Feynman gauge and
$Z_3$ is the wave-function renormalization factor of the gluon
field. Dimensional regularization is used with 
$d=4-2\varepsilon$ being the dimension of space-time.
The Feynman diagrams are divided by
$i{\pi}^{d/2}\exp(-\gamma \varepsilon) {\mu}^{-2 \varepsilon}$
per loop with $\mu$ the \MSb renormalization scale,
$\gamma$ the Euler constant.
The renormalized fermion propagator is obtained by
multiplicative renormalization with a suitable wave-function
renormalization factor $Z_2$
\be
S_{\rm F\,ren}(q) = \frac{1}{Z_2} S_{\rm F}(q)\,.
\ee
We write the inverse propagator in the form
\be
S_{\rm F\, ren}^{-1}(q) = \sla{q}\:B_{\rm ren}-m\:A_{\rm ren}\,.
\lab{Siren}
\ee
Explicitly, we obtain the renormalized amplitudes in the \MSb scheme by
applying the following \MSb renormalization constants~\cite{Tarrach}:
\ba
Z_2 = 1 \z+\z \adp\left(-\xi C_F\right) \frac{1}{\veps}
         +\left(\adp\right)^2 \left(
   \Bigl( \frac{\xi^2}{2} C_F + ( \frac{3\xi}{4} + \frac{\xi^2}{4} ) C_A
   \Bigr) \frac{1}{\varepsilon^2}    \right. \nonumber\\
  \z+\z \left.
   \Bigl( -( \frac{25}{8} + \xi + \frac{ \xi^2}{8} )  C_A
    + t n_f + \frac{3}{4} C_F
   \Bigr )\frac{1}{\varepsilon} \right) C_F,  \nonumber\\
Z_3 = 1 \z+\z \frac{\alpha_s}{8\pi}
  \left( \bigl( \frac{13}{3}-\xi \bigr) C_A - \frac{8}{3}t n_f
     \right) \frac{1}{\veps}  \\
Z_g = 1 \z-\z \adp \left( \frac{11}{3} C_A-\frac{4}{3} t n_f\right)
    \frac{1}{\veps}   \nonumber\\
Z_m = 1 \z-\z \adp \left (3 C_F \right) \frac{1}{\veps}
   +\left( \adp \right)^2 C_F
   \left( (\frac{11}{2} C_A - 2t n_f+\frac{9}{2} C_F)
     \frac{1}{\varepsilon^2} \right. \nonumber\\
     \z-\z  \left.
      (\frac{97}{12} C_A-\frac{5}{3} t n_f+\frac{3}{4} C_F)
       \frac{1}{\varepsilon} \right). \nonumber
\ea
For $SU(N_c)$ we have $C_F=(N_c^2-1)/(2N_c)$,
$C_A=N_c$, $t=1/2$. Since the latter coefficient does not depend
on the number of colors $N_c$ its value will be inserted in most of
the formulae presented below.
\newpage

\section{Renormalized Amplitudes}
There are six diagrams contributing to $\Sigma(q)$ at the two-loop
level (see Fig. 1).\\
\begin{picture}(360,320)(0,0)
%   diagram 1
\Text(40,290)[]{(a)}
\CArc(90,275)(25,180,270)
\CArc(90,275)(25,0,90)
\PhotonArc(90,275)(25,90,180){2}{5}
\PhotonArc(90,275)(25,270,360){2}{5}
\Line(90,250)(90,300)
\Line(50,275)(65,275)
\Line(115,275)(130,275)
\Vertex(90,250){1.8}
\Vertex(90,300){1.8}
\Vertex(65,275){1.8}
\Vertex(115,275){1.8}
%   diagram 2
\Text(220,290)[]{(b)}
\CArc(270,275)(25,180,360)
\BCirc(270,295){15}
\PhotonArc(270,275)(25,130,180){2}{3}
\PhotonArc(270,275)(25,0,50){2}{3}
\Line(230,275)(245,275)
\Line(295,275)(310,275)
\Vertex(245,275){1.8}
\Vertex(295,275){1.8}
\Vertex(255,295){1.8}
\Vertex(285,295){1.8}
%   diagram 3
\Text(40,215)[]{(c)}
\CArc(90,200)(25,180,360)
\PhotonArc(90,200)(25,0,180){2}{10}
\PhotonArc(90,175)(22,32,148){2}{7}
\Line(50,200)(65,200)
\Line(115,200)(130,200)
\Vertex(65,200){1.8}
\Vertex(115,200){1.8}
\Vertex(71,185){1.8}
\Vertex(109,185){1.8}
%    diagram 4
\Text(220,215)[]{(d)}
\CArc(270,200)(25,180,360)
\PhotonArc(270,200)(25,0,180){2}{10}
\Photon(270,175)(270,225){2}{8}
\Line(230,200)(245,200)
\Line(295,200)(315,200)
\Vertex(270,175){1.8}
\Vertex(270,225){1.8}
\Vertex(245,200){1.8}
\Vertex(295,200){1.8}
%     diagram 5
\Text(40,140)[]{(e)}
\CArc(90,125)(25,180,360)
\Line(50,125)(65,125)
\Line(115,125)(130,125)
\PhotonArc(90,125)(25,132,180){2}{3}
\PhotonArc(90,125)(25,0,48){2}{3}
\DashCArc(90,145)(15,0,180){3}
\DashCArc(90,145)(15,180,360){3}
\Vertex(65,125){1.8}
\Vertex(115,125){2}
\Vertex(75,145){1.8}
\Vertex(105,145){1.8}
%     diagram 6
\Text(220,140)[]{(f)}
\CArc(270,125)(25,180,360)
\Line(230,125)(245,125)
\Line(295,125)(310,125)
\PhotonArc(270,125)(25,132,180){2}{3}
\PhotonArc(270,125)(25,0,48){2}{3}
\PhotonArc(270,145)(15,0,180){2}{7}
\PhotonArc(270,145)(15,180,360){2}{7}
\Vertex(245,125){1.8}
\Vertex(295,125){1.8}
\Vertex(255,145){1.8}
\Vertex(285,145){1.8}
%\Text(180,40)[]{Fig.1. Fermion self-energy diagrams to two-loops.}
%\Vertex(0,0){2}
%\Vertex(350,0){2}
%\Vertex(0,400){2}
%\Vertex(350,400){2}
\end{picture}
\vspace*{-3.3cm}

\begin{center} \begin{tabular}{l}
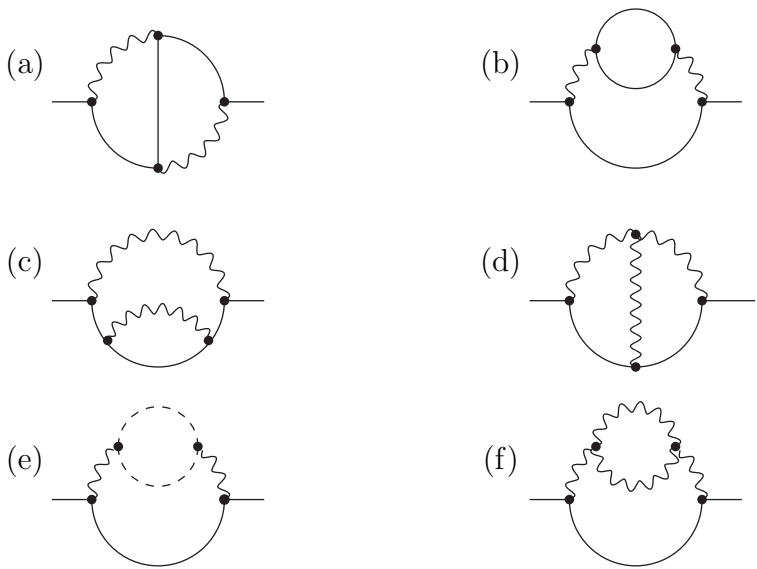
Fig. 1. Fermion self-energy diagrams to two-loops.~~~~~~~~~~~~~~~~~~~\\
Solid lines denote quarks, wavy lines gluons and \\
dashed lines Fadeev--Popov ghosts.	       
	       \end{tabular}
\end{center}

\vspace*{0.5cm}
In our calculation we consider the linear covariant gauge with
arbitrary gauge parameter $\xi$. UV and IR singularities are dealt
with by dimensional regularization and cancel in the observable
quantities. A major problem in performing higher loop calculations
with arbitrary masses is to find an appropriate basis of integrals
which allows us to present results in a compact form. In most cases
integrals can be reduced to a set of master integrals by using partial
fraction decomposition, by differentiation and by integration by
parts. The remaining problem of integrals with irreducible numerators
was solved recently in~\cite{connection}. Such integrals may be
expressed in terms of scalar integrals with dimensions shifted by
multiples of 2 which then are reduced again to integrals of dimension
$d=4-2\varepsilon$ (generic). The resulting complete set of relations
between integrals, a set of generalized recurrence relations, was
considered in~\cite{OVT97} and has been implemented in a computer
program written in FORM~\cite{FORM}.  All diagrams finally could be
expressed in terms of 5 two-loop integrals $I_3$,
$J_{111}(m^2,m^2,m^2)$, $J_{112}(m^2,m^2,m^2)$, $J_{111}(0,0,m^2)$,
$J_{112}(0,0,m^2)$ and 3 products of the one-loop integral structures
$G_{11}(0,m^2)$ and $G_{01}(0,m^2)$, where
\begin{eqnarray}
\z\z I_3(q^2/m^2)=-q^2\int\!\!\!\int \frac{d^dk_1~d^dk_2}
{\pi^d~k_1^2(k_2^2-m^2)((k_1-q)^2-m^2)(k_2-q)^2 ((k_1-k_2)^2-m^2)}\;,
   \non \\
\z\z J_{\alpha \beta \gamma}(m_1^2,m_2^2,m_3^2)=
\int\!\!\!\int \frac{d^dk_1~d^dk_2}{\pi^d~
         (k_1^2-m_1^2)^{\alpha}
        ((k_2-q)^2-m_2^2)^{\beta}
        ((k_1-k_2)^2-m_3^2)^{\gamma}}\;, \non \\
\z\z G_{\alpha \beta}(m_1^2,m_2^2)=\int \frac{d^dk_1}{\pi^{d/2}}\frac{1}
{(k_1^2-m_1^2)^{\alpha} ((k_1-q)^2-m_2^2)^{\beta}}\;.
\label{I3}
\end{eqnarray}
The two-loop integrals $J_{111},J_{112}$ and the one-loop integral $G_{11}$
with zero masses can be expressed in terms of hypergeometric
functions~\cite{BoDa} as follows
\begin{eqnarray}
&&J_{111}(0,0,m^2)=~~i^{2}(m^2)^{d-3}
\Gamma(1-\frac{d}{2})\Gamma(\frac{d}{2}-1)\Gamma(3-d)
\Fh21\Fup{2-\frac{d}{2},3-d}{\frac{d}{2}},\non \\
&& \non \\
&&J_{112}(0,0,m^2)=-i^{2}(m^2)^{d-4}
\Gamma(1-\frac{d}{2})\Gamma(\frac{d}{2}-1)\Gamma(4-d)
\Fh21\Fup{2-\frac{d}{2},4-d}{\frac{d}{2}}, \non \\
&& \non \\
&&G_{11}(0,m^2)=-i(m^2)^{d/2-2}
\Gamma(1-\frac{d}{2})\Fh21\Fup{1,2-\frac{d}{2}}{\frac{d}{2}}.
\label{JJG}
\end{eqnarray}
In appendix A their expansions up to order $O({\varepsilon}^2)$ are given.
$G_{01}$ is elementary:
\bea
G_{01}(0,m^2)=-i(m^2)^{d/2-1}\Gamma(1-\frac{d}{2}).\non
\eea
Finally, $J_{111}$ and $J_{112}$ with equal masses may be written in
the form
\begin{eqnarray}
J_{111}(m^2,m^2,m^2) \z=\z
   m^{2-4\varepsilon}\Gamma^2(1+\varepsilon)
   \left[-\frac{3}{2\varepsilon^2}
  +\frac{1}{\varepsilon}\left( \frac{z}{4}-\frac92 \right)
  +J_3(z)  \right], \nonumber \\
J_{112}(m^2,m^2,m^2) \z=\z
   m^{-4\varepsilon}\Gamma^2(1+\varepsilon)
   \left[ -\frac{1}{2\varepsilon^2}
  -\frac{1}{2\varepsilon}
  +3-\frac{z}{6}+\frac13 J_3(z)-\frac{1}{3} J'_3(z) \right],
\end{eqnarray}
where $z={q^2}/{m^2}$ and $J'_3(z)=zd/d{z}J_3(z)$. The numerical
evaluation of $I_3$, $J_3$ and $J'_3$ will be discussed in appendix B.
Let us mention that results from individual diagrams are rather
lengthy such that we must refrain from presenting individual
contributions.

The amplitudes in (\ref{Siren}) are calculated from the diagrams
of Fig. 1 where diagram (b) was taken with 
$n_f-1$ massless fermions and one massive one in the loop.
We find
%
%        A TERM
%
\begin{eqnarray}
A_{\rm ren} \z=\z
  1 + \adp C_F \left\{4+2\xi + (\xi+3)( \frac{\bz}{z}\lz +L_{\mu}) \right\}
                   \nonumber \\
  \z+\z \left(\adp\right)^2 \left\{
%%%%%%%%%%%%%%%%%%%%%%  CF^2
     C_F^2 \left[
% 27aug99: extrafactor -1/3 here
       -\frac{1}{z}+\frac{260}{3\bz}
% 27aug99: extrafactor -1/3 here without (xi) and factor -1 in terms with xi
      +\frac{103}{3}+\frac{4}{3}z+8\xi+\frac{1+z}{z}\xi^2
        -(1-6\xi+\xi^2)\Bigl(\zeta_2+{\rm Li}_2(z)\Bigr)
                \right. \right. \nonumber \\
       \z-\z \frac{2}{z^2}\left(1-31z+18z^2-\xi z(9-5z)-\xi^2(1-z^2)\right)
          \lz+2(\xi+2)(\xi+3)L_{\mu}
                 \nonumber \\
       \z+\z \frac{2(1+z)}{z} I_3+\frac{8(z+3)}{3\bz}J_3
           -\frac{8(z-9)}{3\bz}{J'_3}
             \nonumber \\
       \z-\z \frac{\bz}{z^3}\left(1-5z+8z^2+6\xi z^2 -\xi^2(1+z)\right)
               \ln^2\bz
                   \nonumber \\
       \z-\z \frac{(\xi+3)}{z}(3z-9-\xi+\xi z)\lz L_{\mu}
               \left. +\frac12(\xi+3)^2 L_{\mu}^2  \right]
               \nonumber \\
%%%%%%%%%%%%%%%%%%%%%%%       CF * CA
       \z+\z C_FC_A \left[
           \frac{1}{24\bz}\left( 235-1291z+16z^2+15\bz\xi(16+3\xi)\right)
            + (7-\frac12\xi+\frac12\xi^2)\Bigl(\zeta_2+{\rm Li}_2(z)\Bigr)
                    \nonumber \right. \\
       \z-\z \frac{(1+z)}{z}I_3-\frac{4(z+3)}{3\bz}J_3
           +\frac{4(z-9)}{3\bz}J'_3
           +\frac{\bz}{z}(\xi+5)\Bigl(\zeta_2\lz+ {\rm Li}_2(z)\,\lz
                    +3S_{1,2}(z)\Bigr) \nonumber \\
        \z+\z \frac{\bz}{12z}(301+60\xi+15\xi^2)\lz
            -\frac{\bz}{2z^2}(1+14z-\bz\xi+z\xi^2) \ln^2\bz
             \nonumber \\
        \z+\z \left. 5\frac{89+12\xi+3\xi^2}{12} L_{\mu}
            +\frac{1}{2z}(\xi^2+3\xi+22)( \bz\lz+\frac{z}{2}
                    L_{\mu}) L_{\mu} \right]
                  \nonumber \\
%%%%%%%%%%%%%%%%%%%%%%%%%%%%%  CF TF
        \z+\z C_F t \left[ 4(1-n_f)\left(\zeta_2+{\rm Li}_2(z)+\frac{8\bz}{3z}
          \lz -\frac{\bz}{z}\ln^2\bz \right) \right.
             \nonumber \\
        \z-\z \frac{52}{3}n_f
          -\frac{32}{3}n_fL_{\mu}-4n_f\frac{\bz}{z}L_{\mu}\lz
              \nonumber \\
        \z-\z 2n_f L_{\mu}^2 + \frac{1}{3\bz^2}(314-95z+15z^2+2z^3)
%>> 27aug99: coeff 1/3 before J3
           -\frac{4(z^2-9)}{3\bz^2}J_3
             \left.\left.  -\frac{4(z-9)}{3\bz}J'_3 \right] \right\}
\end{eqnarray}
%
%    END OF A TERM
%

%\newpage

%%%%%%%%%%%%%%%%%%%%%%
%
%      B TERM
%
\begin{eqnarray}
B_{\rm ren} \z=\z 1+\adp C_F \xi
            \left\{\frac{1+z}{z}+\frac{1-z^2}{z^2}\lz+L_{\mu} \right\}
            \nonumber \\ \z+\z \left(\adp\right)^2 \left\{C_F^2 \left[
%%%%%%%%%%%%%%%%%%%%%%  CF^2
      \frac{1}{z}(4\bz+6z\xi-z\xi^2) \left(\zeta_2+{\rm Li}_2(z)\right)
          \right. \right.  \nonumber \\
    \z+\z \frac{1}{24z}\Bigl(8z^2+183z-1140+48 \xi(8+\xi) \Bigr)
             \nonumber \\
    \z-\z \frac{1}{2z^2}\left(-2(z+3)\bz\xi^2+4(3z-11)\xi
           +11z^2-28z+17 \right) \lz
             \nonumber \\
    \z-\z \frac{4}{z^2}I_3-\frac{2(z+6)}{3z}J_3
           +\frac{2(z-9)}{3z} J'_3
             \nonumber \\
    \z-\z \frac{\bz}{z^3}\left(3+3z-z^2+3\xi(z^2+z-2)
                   -\xi^2 \right) \ln^2\bz
           \nonumber \\
    \z+\z \frac{1}{2z}\left(2\xi^2(1+z)+24\xi-3z\right) L_{\mu}
        +\frac{1}{z^2}(12+\xi-z^2\xi) \xi L_{\mu}\,\lz
              +\left. \frac12 \xi^2 L_{\mu}^2 \right]
           \nonumber \\
%%%%%%%%%%%%%%%%%%%%%%%%%  CA * CF
    \z+\z C_FC_A \left[
         \frac{1}{2z}(4+2z-\xi z+z\xi^2)\Bigl(\zeta_2+{\rm Li}_2(z)\Bigr)
         -\frac{1}{24z}\Bigl(4z^2-147z-714
                \nonumber \right. \\
     \z-\z 12(7+13z)\xi - 9(2+3z)\xi^2\Bigr)
          +\frac{\bz}{z^2}(z+4+\xi z)\Bigl(\zeta_2 \lz
             +{\rm Li}_2(z)\,\lz + 3S_{1,2}(z) \Bigr)
                  \nonumber \\
     \z+\z \frac{2}{z^2}I_3+\frac{z+6}{3z} J_3
          -\frac{(z-9)}{3z}J'_3
         +\frac{\bz}{4z^2}\Bigl(11z+35+14\xi(1+z)+3\xi^2(1+z)\Bigr)\lz
               \nonumber \\
    \z-\z \frac{\bz}{2z^2}(1+\xi)\Bigl(1+\xi(1+z)\Bigr) \ln^2\bz
             +\frac{1}{4z} \left(\xi^2(2+3z)
                  +2\xi(3+7z)+25z \right) L_{\mu}
                         \nonumber \\
    \z+\z \left. \frac{(1-z^2)}{2z^2} \xi(3+\xi) L_{\mu}\,\lz
            +\frac14 \xi(3+\xi) L_{\mu}^2
              \right] \nonumber \\
%%%%%%%%%%%%%%%%%%%%  CF * TF
    \z+\z C_F t \left[ \frac{236}{5\bz^2}+\frac{11}{15\bz}
      -\frac{9z^2-120z-56}{30z}
       +\frac{ 8(13-z) }{15\bz^2}J_3
          -\frac{4(z-2)(z-9)}{15z\bz}J'_3 \right.
          \nonumber \\
    \z-\z  \left. \left. \frac{1}{2z}(7z+4)n_f
       -2n_fL_{\mu}- 2\frac{\bz}{z^2}
            \Bigl((1+z)n_f+\frac{1}{15}(z^2-17z-14)\Bigr)
              \lz  \right] \right\},
% \right\}
\end{eqnarray}
%
%    END OF B TERM
%
where $\bz = 1-z$ and $ L_{\mu}=\ln(\mu^2/m^2)$. The polylogarithms
${\rm Li}_n(z)$ and $S_{n,p}(z)$ are defined by the integrals
\begin{eqnarray*}
S_{n,p}(z)=\frac{(-1)^{n+p-1}}{(n-1)!\: p!}\int_0^1 
\frac{\ln^{n-1}(x)\:\ln^p(1-xz)}{x}\:dx \;\;;\;{\rm Li}_n(z) = S_{n-1,1}(z)
\end{eqnarray*}
and $\zeta_n=\zeta(n)=\sum_{i=1}^{\infty}1/n^i$ is the Riemann zeta
function, with values $\zeta_2=\pi^2/6$, $\zeta_3=1.202057...$, 
$\zeta_4=\pi^4/90$ etc..

We observe that in order to evaluate the on-shell values for $A_{\rm
 ren}$ and $B_{\rm ren}$ ($\bz \to 0$) we need to know the on-shell
 value of $I_3(z)$ and the expansions of $J_3(z)$ up to order
 ${\bz}^2$ and of $J_3'(z)$ up to order $\bz$.  The on-shell value of
 $I_3(z)$ (at $z=1$) is known~\cite{Br} and is given for completeness
 in appendix B. The expansions of the integrals $J_{111}$ and
 $J_{112}$, and the corresponding expansions of $J_3(z)$ and
 $J_3'(z)$, can be obtained by the second-order differential equation
\begin{eqnarray*}
&&2(q^2+m^2)(q^2+9m^2)m^2J_3^{\prime\prime}
-\left((d-4)q^4+10(3d-10)q^2m^2+9(5d-16)m^4\right)J_3^\prime
\nonumber\\&&{}
+3(3d-8)(d-3)(q^2+3m^2)J_3
=\frac{48q^2m^{2d-6}}{(d-4)^2}\,,
\end{eqnarray*}
where primes denote differentiation w.r.t.\ $m^2$, which was derived
in~\cite{BFT}. As a result we find
\bea
&&J_{3}(z)=-\frac{59}{8}+\frac{3}{8} \bz+
(\frac{5}{4}-\frac{3}{4} {\zeta}_2) {\bz}^2, \non \\
&& \non \\
&&J_{3}'(z)=-\frac{3}{8} - (\frac{17}{8}-\frac{3}{2} {\zeta}_2)\,.
\eea

  In appendices D and E we also present the small and large momentum
expansions of the renormalized amplitudes $A_{\rm ren}$ and $B_{\rm
ren}$. For details concerning the expansions of $I_3(z)$ and $J_3(z)$
we refer to appendix B. 

%\newpage

\section{Connection between pole mass and \MSb mass}
Due to the confinement of QCD, quark masses in general do not have an
unambiguous simple physical meaning. This is especially true for the
masses of the light quarks which are usually parametrized by \MSb
masses, the \MSb renormalized Lagrangian mass parameters. Physical
values may be attached to them as current quark masses which cause the
observed chiral symmetry breaking (see
e.g.~\cite{GasserLeutwyler}). For a very heavy quark, like the top
quark, the situation is different. The top quark is so unstable that
it decays by weak interactions before strong interactions come into
play and form bound states. In this case the complex pole mass gains a
physical meaning as it manifests itself in a resonance peak in
physical cross sections. In the following we calculate the pole mass at
the two-loop level in perturbative QCD. We should mention here
that the true pole mass of full QCD is expected to differ by
non-perturbative effects from the pole mass obtained in perturbative
QCD.

The relationship between the pole mass and the \MSb mass at two-loop
order was first obtained in~\cite{GBGS}. The two-loop correction turns
out to be rather large. This may have physical reasons~\cite{BLM}, but
there could also be a problem with the procedure of calculation
adopted in~\cite{GBGS}. In~\cite{GBGS} prior to the
$\varepsilon$-expansion and the renormalization, the external momentum
was taken to be on-shell and recurrence relations were applied
directly to the on-shell integrals. Note that the corresponding
interchange of integrations with taking the on-shell limit could cause
problems due to possible infrared singularities. One expects that such
an on-shell algorithm can be applied for infrared-safe
quantities. However, since we are lacking a rigorous proof that the
recurrence relations for the on-shell integrals yield the correct
answer, it is highly desirable to find the connection between the pole
mass and the \MSb mass by means of a different method which avoids
possible ambiguities. This is possible by determining the position of
the pole of the \MSb renormalized propagator. Thus, setting
$\sla{q}=M$ in
\re{Siren}, and taking the limit $q^2 \rightarrow M^2$ in $A_{\rm
ren}$ and $B_{\rm ren}$, we obtain the transcendental equation
\begin{equation}
\lim_{q^2 \to M^2} [m A_{\rm ren}(q^2,m^2) - M B_{\rm ren}(q^2,m^2)]=0
\end{equation}
for the pole mass $M$. Its iterative solution yields the desired
relation
\begin{equation}
M=m \left( 1+c_1 \left( \frac{\alpha_s}{4 \pi} \right)
     +c_2 \left( \frac{\alpha_s}{4 \pi} \right)^2 + \ldots \right),
\lab{Mm}
\end{equation}
with
\begin{eqnarray}
&&c_1=C_F(4+3L), \\
&& \nonumber \\
&&c_2=
 C_FC_A \left(\frac{1111}{24}-8\zeta_2 - 4I_3(1)+\frac{185}{6}L
 +\frac{11}{2}L^2\right)
  \non \\
&& \non \\
&&~~~
 -C_F t n_f \left(\frac{71}{6}+8\zeta_2+\frac{26}{3}L+2L^2 \right)
  \non \\
&& \non \\
&&~~~
+C_F^2\left(\frac{121}{8} +30\zeta_2 + 8I_3(1)
 +\frac{27}{2} L+\frac92 L^2 \right)-12C_F t \left(1-2\zeta_2 \right).
\end{eqnarray}
where $I_3(1)$ is given in appendix B and $L=\ln(\mu^2/M^2)$.
Setting $\mu^2=M^2$ we arrive at the relation already known
from~\cite{GBGS}.

   Note that as long as we consider QCD corrections only, the location
of the pole of the quark propagator $M$ is real. Only after switching
on the weak interaction the quarks become unstable and the pole
moves off the real axis.

   Our result confirms the validity of the on-shell approach, i.e. the
possibility of setting the external momentum squared on-shell and
applying the on-shell recurrence relations from the very
beginning. This check thus enhances our confidence that it is possible
to derive the same relation on the three-loop level by starting from
the on-shell recurrence relations, which for this task evidently is
much simpler than its ``indirect'' evaluation from the renormalized
propagator.

  After completion of this paper we learned about related
work~\cite{Franco} where the 2-loop relation between the quark mass in
the $\overline{\rm MS}$ and MOM schemes has been obtained in the NNLO
approximation. The result~\cite{Franco} corresponds to the limit
$q^2/m^2\to\infty$ of the complete 2-loop massive quark propagator
which has been presented here.\\

{\large \bf Acknowledgments}\\

  This work has been supported in part by the BMBF (O.V.T and O.L.V.).
O.V.T. and O.L.V. are grateful to the Physics Department of the
University of Bielefeld for its warm hospitality. We thank R. Scharf
and V. Lubicz for pointing out two misprints in our preprint
hep-ph/9803493.\\
\smallskip

\newpage

{\Large\bf Appendix A}\\

Here the two-loop integrals $J_{111}(0,0,m^2), J_{112}(0,0,m^2)$ and
$G_{11}$ of (\ref{JJG}) are expanded in $\varepsilon$. This is the form
which is needed to obtain the renormalized amplitudes $A$ and $B$.
\small
\begin{eqnarray}
&&e^{2\gamma \varepsilon} 
    m^{4\varepsilon-2}J_{111}(0,0,m^2)= -\frac{1}{2 \varepsilon^2}
    +\frac{z-6}{4\varepsilon} - 3 + \frac{13}{8}z
    - \frac{3}{2} \zeta_2+\frac{(1-z^2)}{2z} \ln\bz
    -\Li2(z) \nonumber \\
&&~~+ \varepsilon  \left(  - \frac{15}{4}+ \frac{115}{16} z
 +\frac{3(z-6)}{4} \zeta_2 + \frac43 \zeta_3  + \frac{1-z^2}{z}
   \left( \frac{13}{4} - \ln\bz \right) \ln\bz \right.
\nonumber \\
&&~~\left.
 + \frac{2z^2-6z-1}{2z} \Li2(z)
 - \Li3(z)-4 S_{1,2}(z) \right)
 +\varepsilon^2 \left( \frac{21}{8}+\frac{865}{32}z -9\zeta_2+
 \frac{39}{8}z\zeta_2
 \right.
 \nonumber \\
&&~~  +(4-\frac23 z) \zeta_3-\frac{63}{8}\zeta_4
+\frac{1-z^2}{z} \left( \frac{115}{8}+\frac32\zeta_2
- \frac{13}{2} \ln\bz
 +\frac{4}{3} \ln^2\bz \right) \ln\bz
  \nonumber \\
&&~~
  - 3 \Li2^2(z)- \Li4(z)
 -\frac{2}{z}(z^2+6z-2) S_{1,2}(z)+\frac{1}{2z}(2z^2-6z-1) \Li3(z)
\nonumber \\
&& ~~ \left.
 -( 6 + \frac{13}{4z} -\frac{13}{2} z +3\zeta_2) \Li2(z)
 +\frac{3(1-z^2)}{z} \ln\bz \Li2(z)
 -16 S_{1,3}(z)+8 S_{2,2}(z) \right),  \nonumber \\
\end{eqnarray}
\normalsize
where $S_{n,p}(z)$ is the generalized Nielsen polylogarithm~\cite{DD}.
\small
\begin{eqnarray}
&&e^{2\gamma \varepsilon} m^{4\varepsilon-4}J_{112}(0,0,m^2)=
-\frac{1}{2\varepsilon^2}
 -\frac{1}{2\varepsilon} +\frac12 -\frac{3}{2}\zeta_2
 +\frac{\bz}{z} \ln\bz - \Li2(z)
\nonumber \\
&&~~ +\varepsilon  \left( \frac{11}{2} -\frac{3}{2}\zeta_2
 + \frac{5\bz}{z} \ln\bz - \frac{2\bz}{z} \ln^2\bz
 + \frac43 \zeta_3  - \frac{1}{z} \Li2(z)  - \Li3(z) \right.
 \nonumber \\
&&~~\left. -4 S_{1,2}(z) \right)+\varepsilon^2 \left( \frac{49}{2}
 + \frac32 \zeta_2 - \frac{63}{8} \zeta_4 +\frac{\bz}{z}
 (19 + 3 \zeta_2) \ln\bz - 10\frac{\bz}{z}\ln^2\bz
 \right.
 \nonumber \\
&&~~  +\frac{8\bz}{3z} \ln^3\bz + \frac43 \zeta_3
       -\frac{1}{z} \Li3(z)
 + \Li2(z)  ( 6 -\frac{5}{z} -3 \zeta_2 +6\frac{\bz}{z} \ln\bz )
 \nonumber \\
&&~~ \left.
 + (  \frac{8}{z} - 12  ) S_{1,2}(z)
       - 3 \Li2^2(z) - \Li4(z)
       -16 S_{1,3}(z) + 8 S_{2,2}(z)
       \right).
\end{eqnarray}
%
% id mas2(0,mm,1)/c1/c2*td1(1,mm,0)=
%
\begin{eqnarray}
&&e^{2\gamma \varepsilon} m^{4\varepsilon-2}G_{11}(0,m^2)G_{01}(0,m^2)=
-\frac{1}{\varepsilon^2}
  -\frac{1}{\varepsilon}
  \left(3+\frac{\bz}{z} \lz \right) - 7 -\zeta_2
  + \frac{\bz}{z} \Bigl( - 3\lz + \ln^2\bz + \Li2(z) \Bigr)
  \nonumber \\
&&~~
+\varepsilon
  \Biggl[ - 15 -3\zeta_2+ \frac23 \zeta_3+\frac{\bz}{z} \Bigl(
  -7 - \zeta_2 +3 \lz - \frac{2}{3} \ln^2\bz \Bigr) \lz
       \nonumber \\
&&~~
 + \frac{\bz}{z}\Biggl(
  \Bigl( 3 - 2 \lz \Bigr) \Li2(z)
       + \Li3(z)-2 S_{1,2}(z) \Biggr)
  \Biggr]
  \nonumber \\
&&~~+\varepsilon^2 \frac{\bz}{z}
 \Biggl[  \frac{z}{\bz} \Bigl(  - 31 - 7 \zeta_2 +2\zeta_3
        - \frac74 \zeta_4 \Bigr)
  - \left(15  + 3  \zeta_2
  - \frac{2}{3} \zeta_3 \right)\lz
      \nonumber \\
&&~~
  + (7+\zeta_2) \ln^2\bz
  - 2 \ln^3\bz + \frac{1}{3}\ln^4\bz
  + \left( 7 + \zeta_2 - 6 \ln\bz + 2 \ln^2\bz  \right) \Li2(z)
 \nonumber \\
&&~~
   + \Bigl( 3 - 2 \ln\bz \Bigr) \Bigl( \Li3(z)-2S_{1,2}(z)  \Bigr)
   +\Li4(z) + 4S_{1,3}(z)-2 S_{2,2}(z)
 \Biggr].
\end{eqnarray}
\normalsize
The last formula can as well be used for the $\varepsilon$-expansion
of the one-loop scalar propagator integral with arbitrary
masses and momentum (see, for example, \cite{bds2}).

A surprising fact is that in the final renormalized result of the
quark propagator given in sect.~3 various structures like $\zeta_3,
\zeta_4, \Li2^2(x), \Li3(x), \Li4(x)$ and all generalized
Nielsen polylogarithms, except $S_{1,2}$, have canceled.\\

\smallskip

{\Large \bf Appendix B}\\

   In this appendix we present details about the calculation of the
functions $I_3(z)$ and $J_3(z)$ entering the self-energy function
given in sect.~3. We can give one-dimensional integral representations
which are, however, not the most convenient forms for a numerical
evaluation as will be seen. In particular it turns out that the
expansions w.r.t. small and large $q^2$ are very efficient even in the
time-like region (on the cut) of the functions. For this reason we
give a full account of these expansions, mainly due to~\cite{Br}.

   An integral representation for $I_3$ is given by the following
expression~\cite{Br,BFT}
\begin{eqnarray}
&&q^2I_3(q^2/m^2)=\int_m^{\infty}\frac{4w~dw}{w^2-m^2}\left( \ln\frac{w}{m}
 -\frac{w^2-m^2}{w^2}\ln \frac{w^2-m^2}{m^2} \right)
 \ln \frac{w^2+q^2}{w^2} \nonumber \\
&&\nonumber \\
&&~~ +\int^{\infty}_{2m} \frac{4~dw}{(w^2-4m^2)^{1/2}}
 \left(3 \ln \frac{w}{m} - \frac{3w^2-3m^2+q^2}{W_+W_-}
 \ln \frac{W_++W_-}{W_+-W_-}\right) \ln \frac{w}{2m},
\label{I3int}
\end{eqnarray}
with $W_{\pm} \equiv ((w\pm m)^2+q^2)^{1/2}$.\\

   We recall that for $q^2=m^2$ ($z\equiv q^2/m^2=1$) the value
of $I_3$ is~\cite{Br}

\begin{equation}
I_3(1)=\frac{3}{2} {\zeta}_3 - 6{\zeta}_2 \ln 2.
\end{equation}

The small $q^2$ expansion is obtained from~\cite{Br}
in the following manner:
with the Ansatz
\begin{equation}\label{ansatzI3}
I_3(z)=-\sum_{n=1}^{\infty}z^n\frac{d_3(n)}{n},
\end{equation}
$d_3(n)$ being expansion coefficients and
$ d_3(1)=-{\zeta}_2+\frac{27}{2} S_2 $ with $S_2$ derived from the maximum
value of Clausen's integral,
\begin{equation}
S_2 = \frac{4}{9 \sqrt{3}}{\rm Cl}(\frac{\pi}{3}) = 0.2604341376 \ldots~~~~.
\end{equation}

   The higher order terms are obtained by solving the following differential
equation~\cite{Br} (the notation is adopted from~\cite{Br}) 
\begin{equation}
D_3 D_2 I_3(z) = D_3 D_2 I_a(z) + (q^2-9 m^2)^2-2{\pi}^2 m^2 (q^2+3m^2),
\label{Deq}
\end{equation}
where
\bea
&&D_2 \equiv (q^2-m^2)\frac{d}{d q^2},
\nonumber \\
&&\nonumber \\
&&D_3 \equiv 6 m^2 (q^2+3m^2) + (q^2-9 m^2) \left[
(q^2-m^2)(q^2-9 m^2)\frac{d}{d q^2} -8 m^2 \right]q^2 \frac{d}{d q^2}
\eea
and
\begin{equation}
I_a(z)=\sum_{n=1}^{\infty}z^n \left\{
-\frac{\pi^2}{6n}+\sum_{r=1}^{n-1}(\frac{1}{r^2 n}-\frac{2}{r n^2}) \right\}.
\end{equation}

Inserting (\ref{ansatzI3}) into (\ref{Deq}), equating equal
powers  of $z$ and solving  iteratively for the coefficients $d_3(n)$
is very conveniently
done, e.g., with the help of a REDUCE~\cite{RED} program, which can also
perform the final numerical evaluation. To calculate $I_3$ on its cut,
the method of mapping and Pad\'{e} approximation is applied as
introduced in~\cite{FT}. Since any number of Taylor coefficients is
easily obtained for $I_3$, it can be calculated with practically any
required precision even on the cut. Thus as an example for z=100 with
100 Taylor coefficients we obtain $I_3=7.77982600-0.77916795i$, where
the precision is estimated from the convergence of the Pad\'{e}
approximants. For comparison: the integral representation
(\ref{I3int}) is not even applicable on the cut and in the space-like
region for such large $q^2$ the two contributions mainly cancel, which
seems not the proper way to evaluate $I_3$ numerically.

   As described in~\cite{Br} as well, for large $q^2$ the large momentum
expansion can be used: with five terms
\bea
&&{(I_3)}_{\rm asy}(z)=6\zeta_3+(2L_z^2+6L_z+6)/z
 +(4L_z^2+\frac{1}{2}L_z-\frac{15}{4})/z^2+ \nonumber\\
&&(\frac{29}{3}L_z^2-\frac{46}{3}L_z-\frac{257}{18})/z^3+
  (32L_z^2-\frac{1957}{24}L_z-\frac{3613}{96})/z^4+ \nonumber\\
&&\nonumber \\
&&(\frac{677}{5}L_z^2-\frac{30907}{75}L_z-\frac{103577}{1000})/z^5\,,
\eea
and $L_z=\ln z-i\pi$ we obtain the same numerical result
as above except for the
last decimal in the imaginary part
(i.e. ${\rm Im}~{(I_3)}_{\rm asy}(100)=-0.77916794$).

We also mention that from the small
and large momentum expansions of $I_3(z)$
a 3-loop integral $D_3$ was calculated~\cite{RT}.

The situation is similar for $J_3$.
We found the following integral representation
\begin{eqnarray}
J_3(z) \z=\z
  -9+\frac{13}{8}z + \frac{(3+z)(1-z)}{2z}\ln(1-z) \nonumber\\
  \z\z
  +(1-z)\int_0^1\frac{dy [-3+y(3-z)-3y^2]}{2zy(1-y)^2}
  \left( 1-\frac{1-y+y^2-zy}{\sqrt{\Delta}} \right) \ln y,
\end{eqnarray}
with $\Delta=y^4-2(1-z)y^3+(z^2-6z+3)y^2-2(1-z)y+1$. However, this
representation is only applicable for $z \leq$ 1~. Recall that the
first threshold starts at $z$=9 ! Therefore, the small and large
momentum expansions will be much more useful again. For $J_3(z)$ we
have the special values~\cite{spev}
\bea
J_3(1)=-\frac{59}{8}\;,\;\;\; J_3(9)= \frac{45}{8}-\frac{8
\pi}{\sqrt{3}}\;.
\eea
Writing
\begin{equation}
J_3(z)=\sum_{n=0}^{\infty}z^n j_3(n),
\end{equation}
the Taylor coefficients $j_3(n)$ of $J_3$ are obtained recursively in
the following manner
\begin{equation}
 j_3(0)=-\frac{21}{2} +\frac{27}{2}S_2 ,~j_3(1)= \frac{ 3}{ 8}- 3 S_2,
~j_3(2)= \frac{11}{108}- \frac{1}{3}S_2 ,~j_3(3)= \frac{19}{648}-
\frac{1}{9}S_2.
\end{equation}
\noindent
Introducing $c_3(j)=-j!(j+1)!j_3(j), j \geq 2$, we have the
recursion~\cite{BFT}
\begin{equation}
c_3(j+1)=\frac{1}{9} \left( (10 j^2-4) c_3(j )-(j-2) (j-1) j (j+1)
c_3(j-1) \right).
\end{equation}
Thus the coefficients are much easier to calculate than in the case of
$I_3$.  Furthermore, the precision of the Pad\'{e} approximants is
much higher, i.e. already with 33 coefficients one obtains for $z=100$
a precision of 10 decimals. The reason for this is of course that the
threshold is much higher in this case.

The large $q^2$ expansion of $J_3$ is given in~\cite{BFT}.
The first six terms
(again up to $z^{-5}$) give the contribution
\begin{eqnarray}
&&\left(J_3 \right)_{\rm asy}(z)=
 -(\frac12 L_z-\frac{13}{8}) z
 +\frac32 L_z^2-\frac{15}{2}
 -(3 L_z^2+\frac92 L_z-\frac34)/z             \nonumber \\
&&~~ -(3 L_z^2-6 L_z-\frac{29}{4})/z^2
     -(9 L_z^2-23 L_z-\frac{307}{24})/z^3
     -(36 L_z^2-107 L_z-\frac{7927}{240})/z^4 \nonumber \\
&&~~
     -(171 L_z^2-\frac{2773}{5} L_z-\frac{14107}{150})/z^5.
\end{eqnarray}
For $z$=100 this yields also a precision of 10 decimals, i.e. the same
precision as the small $q^2$ expansion with 33 coefficients.

What concerns $J_3'$, the Taylor series as well as the large momentum
expansion are easily obtained by differentiation from the
corresponding representations of $J_3$.  The precision with the same
number of coefficients as above is here 9 decimals for the small $q^2$
expansion and the asymptotic expansion.

Finally we mention that also the Taylor coefficients can be calculated
analytically and stored which allows us to write a FORTRAN program in
terms of the multiple precision version of
D.H. Bailey~\cite{Bail}. Then, calculating these functions from the
small momentum expansion, for $I_3$, $J_3$ and $J_3'$ only the
Pad\'{e}'s have to be performed, which is very little time consuming.
For the numerical evaluation of $S_{1,2}$ we refer the reader
to~\cite{DD}.\\

%\newpage

{\Large \bf Appendix C}\\

   In the following we present the bare amplitudes $A_{\rm bare}$ and
$B_{\rm bare}$ in terms of the basic integrals of sect.~3. For these we
introduce the following notation:
\bea
%
% V(1,1,1,1,0)
&&I_1=G_{11}(0,m^2) G_{11}(0,m^2),\\
%
%  mas2(0,mm,1) td1(1,mm,0) c1^(-1) c2^(-1)
&&I_2=\frac{1}{m^2}G_{1,1}(0,m^2) G_{01}(0,m^2),\\
%
%       FM(0,mm,mm,0,mm)
&&I_3~~{\rm see}~~ (\ref{I3}), \\
%
%    mas3(mm,mm,mm) c1^(-1) c2^(-1) c3^(-1)
&&I_4=\frac{1}{m^2}J_{111}(m^2,m^2,m^2), \\
&&I_5=J_{112}(m^2,m^2,m^2), \\
%
% mas3(0,0,mm) c1^(-1) c2^(-1) c3^(-1)
&&I_6=\frac{1}{m^2}J_{111}(0,0,m^2), \\
%
% mas3(0,0,mm) c1^(-1) c2^(-1) c3^(-2)
&&I_7=J_{112}(0,0,m^2), \\
%
% td1(1,mm,0) td1(1,mm,0)
&&I_8=\frac{1}{m^4}G_{01}(0,m^2) G_{01}(0,m^2)
\eea

   The general structure of the bare amplitudes reads
%textAbare:=
\begin{eqnarray}
A\z=\z
  \frac{C_F}{1-z} \left( \frac{C_Aa_1}{d-4}+\frac{C_Fa_2}{z} \right) I_1
   + \frac{C_F}{1-z} \left( \bigl( C_F - \frac12 C_A \bigr) a_{3}
       + \frac{t a_{4}(3 d-8)}{3(1-z)^2} \right) I_4
\nonumber \\
  \z+\z \frac{(d-2)  C_F}{(1-z)(d-3)} \left(\frac{C_A a_5}{d-4}
       + \frac{(1-z)(d-3)t}{d-2} a_6
       + \frac{ C_Fa_7}{z}
       \right)I_2
 \nonumber \\
 \z+\z  \frac{C_F}{z} \left( C_F - \frac12 C_A \right) a_8 I_3
       + \frac{C_F}{1-z} \left( \bigl( C_F  - \frac12 C_A \bigr) a_{9}
       + \frac{t a_{10}}{3(1-z)^2} \right) I_5
 \nonumber \\
 \z+\z  \frac{(3 d-8)C_F}{(d-4)(1-z)}
        \left( \frac{C_A a_{11} }{2 (d-4)}
       +\frac{t a_{12}}{3 d-8} + C_F a_{13} \right) I_6
  \nonumber \\
  \z+\z \frac{C_F}{(d-4)(1-z)}
        \left( \frac{C_A  a_{14}}{(d-4)(d-3)}
       + t a_{15}
       + \frac{C_F a_{16}}{d-3} \right) I_7
  \nonumber \\
  \z+\z  \frac{(d-2)^2 C_F}{(1-z)(d-3)}
       \left(\frac{C_A a_{17}}{d-4}
       + \frac{ t  a_{18}}{3 (1-z)^2 (d-5)}
       + \frac{C_F a_{19}}{z}\right )I_8
\end{eqnarray}

%  valueBbare:=
\begin{eqnarray}
zB \z=\z
   \frac{C_F}{1-z} \left( \frac{C_A b_1}{d-4}
       + C_F b_2 \right) I_1
     +  \frac{(d-2) C_F}{(d-3)(1-z)}
        \left(\frac{  C_A b_3 }{d-4}
       +  \frac{t(1-z)(d-3) b_4}{15(d-2)} \right.
 \nonumber \\
 \z+\z   C_F b_5 \Bigr)I_2
      + \frac{C_F}{z} \left( C_F - \frac12 C_A \right)  b_6 I_3
     + \frac{C_F}{1-z} \left( \bigl( C_F - \frac12 C_A \bigr) b_{7}
     +  \frac{t b_{8} }{15(1-z)^2} \right) I_4
  \nonumber \\
  \z+\z   \frac{C_F}{1-z} \left( \bigl( C_F - \frac12 C_A \bigr) b_9
      +  \frac{t b_{10}}{15 (1-z)^2} \right) I_5
 +  \frac{ C_F}{1-z} \left( \frac{C_Ab_{11}}{2(d-4)^2(d-6)}
       + \frac{t b_{12}}{d-6}
       + \frac{ C_F b_{13}}{d-4} \right) I_6
  \nonumber \\
  \z+\z  \frac{C_F}{1-z} \left(\frac{ C_A b_{14}}
        {(d-3)(d-4)^2(d-6)} + \frac{ t  b_{15}}{d-6}
        + \frac{C_F b_{16}}{(d-4)(d-3)} \right) I_7
  \nonumber \\
  \z+\z  \frac{(d-2)C_F}{(d-3)(1-z)}
        \left(\frac{C_A  b_{17} }{d-4}
       + \frac{ t b_{18} }{15 (d-5)  (1-z)^2}
       + C_F b_{19} (d-2) \right)I_8,
\end{eqnarray}
where the coefficients $a_i$ and $b_i$ (~i=1,2, \ldots ~) are polynomials
in the space-time dimension $d$, the gauge parameter $\xi$ and
$z=q^2/m^2$.

\small
\begin{eqnarray*}
a_1 &=& (d-4)(d^2-4d+2)-(d-2)(d^2-18 d+46) z-2(d-3)(d-4)
      \xi-2(d-3)(d-6)\xi z\\
%\\
a_2 &=& (d-2)^2+(-4-4 d^2+12 d) z+(12-d^2) z^2+4 (d-1) (d-4) \xi
z+4 (d-1) (d-2) \xi z^2 \\
& & - (d-2) (d-6) \xi^2 z^2+(-2 d^2-20+12 d)
\xi^2 z-(d-2)^2 \xi^2\\
%\\
a_{3} &=& 8 d (3 d-8)\\
%\\
a_{4} &=& -48 d+264+(-96 d+272) z+(16 d-24) z^2\\
%\\
a_5 &=& -d^3-86+47 d-3 d^2+(2+d^3-9 d^2+17 d) z+(3 d-10) (d-5)
\xi+(d-2) (d-5) \xi z\\
%\\
a_6 &=& 8/3 (d-1) (d-2)\\
%\\
a_7 &=& -2 (d-2) (d-3)-2 d (-13 d+22+2 d^2) z+(-2 d+12) z^2\\
& & -2(d-1)(5 d-16) \xi z-2 (d-1) (d-2) \xi z^2+2 (d-3) (d-4) \xi^2 z+2
 (d-2) (d-3) \xi^2\\
%\\
a_8 &=& 24+2 d^2-12 d-(2 d^2+8-12 d) z\\
%\\
a_{9} &=& 8 z d-72 d\\
%\\
a_{10} &=& -2160+432 d+(-2352+816 d) z+(432-240 d) z^2+(-16+16 d)
z^3\\
%\\
a_{11} &=& -(d-4) (d^2-9 d+16) \xi^2+(2 d^3-8-18 d^2+40 d) \xi-d^3-
7 d^2+60 d-104\\
%\\
a_{12} &=& -8 (d-2) (3 d-8) (n_f-1)\\
%\\
a_{13} &=& 2 (d-3) (d-6) \xi^2-8 (d-1) (d-4) \xi-4-14 d+6 d^2\\
%\\
a_{14} &=& -216 d^2+628 d+26 d^3-640+2 (d-2) (d-4) (d^2-3 d+1) z+
(332 d-4 d^4+50 d^3\\
& &- 128 - 210 d^2) \xi-2 (d-5) (d-2) (d-4) \xi z+2
 (d-4) (d-3) (d^2-9 d+16) \xi^2\\
%\\
a_{15} &=& 32 (n_f-1) (d-2)\\
%\\
a_{16} &=& -144+68 d^2-16 d-16 d^3-4 (d-2) (d-4) (2 d-3) z \\
& & +4 (7d-22) (d-1) (d-4) \xi+4 (d-1) (d-2) (d-4) \xi z-8 (d-6) (d-3)^2
\xi^2\\
%\\
a_{17} &=& (-d+5) \xi+(d+1) (2 d-7)\\
%\\
a_{18} &=& -8 d^2+204-4 d+(80 d^2-616 d+1112) z-4 (d-1) (2 d-9) z
^2\\
%\\
a_{19} &=& d-3+(2 d^2-10 d+4) z+(4 d-4) \xi z+(3-d) \xi^2
\end{eqnarray*}

\vspace{0.5cm}

\begin{eqnarray*}
b_1 &=& -2 (d-2) (d-3) \xi z^2-2 (d-3) (d-4) \xi z+(d-2) (d^2-10
d+22) z^2\\
& & +(56-34 d+8 d^2-d^3) z\\
%\\
b_2 &=& -3 (d-2)^2+(36-12 d) z-(d-2) (d-6) z^2+4 (d-1) (d-2) \xi
+4 (d-1) (d-4) \xi z \\
& & -(d-2)^2 \xi^2 z^2+(-2 d^2-20+12 d) \xi^2 z-(
d-2) (d-6) \xi^2\\
%\\
b_3 &=& -56+24 d-2 d^2+(d^3+6-5 d^2+5 d) z-(d-2) (d^2-9 d+19) z
^2 \\
& & +(d-2) (d-4) \xi+(d-3) (3 d-10) \xi z+(d-2) \xi z^2\\
%\\
b_4 &=& 2 (d-2) (d-4)-4 (d-2) (d-4) z+2 (d-2) (d-4) z^2\\
%\\
b_5 &=& -12-10 d+4 d^2+(-20+16 d-2 d^2) z+2 (d-2) (d-5) z^2 \\
& & +(d-
1) (2 d^2-15 d+26)\xi+2 (d-1) (d-3) (d-4) \xi z+(d-1) (d-2) \xi
z^2\\
& & +2 (d-2) (d-3) \xi^2 z+2 (d-3)(d-4) \xi^2\\
%\\
b_6 &=& -2 d (d-6) + 2 (d-2) (d-4) z\\
%\\
b_{7} &=& -2 (7 d-18) (d-6)-10 (d-2)^2 z\\
%\\
b_{8} &=& 36 (7 d-20) (d-3)+(-4680 d+576 d^2+8336) z+(2288+408 d
^2-1928 d) z^2\\
& & +(472 d-592-96 d^2) z^3+4 (3 d-8) (d-3) z^4\\
%\\
b_9 &=& -216+36 d+(32 d-48) z+(-4 d+8) z^2\\
%\\
b_{10} &=& -756 d+2160+(8832-1644 d) z+(4 d-16) z^5+(2160-1032 d)
z^2 \\
& & +(-1072+424 d) z^3+(224-68 d) z^4\\
%\\
b_{11} &=& -3168 d-8 d^5+2624 d^2+132 d^4-874 d^3+768-(d-2) (d^4+
5 d^3-92 d^2+312 d\\
& & -320) z+4 (3 d-8) (d-4) (d^3-9 d^2+25 d-18)
 \xi \\
 & & +2 (3 d-8) (d-2) (d^3-7 d^2+4 d+28) \xi z-(3 d-8) (d-2) (d
-4) (d^2-9 d+16) \xi^2 z \\
& & -2 (3 d-8) (d-3) (d-4)(d^2-9 d+16) \xi^2\\
%\\
b_{12} &=& 16 (d-2) (d-3) (n_f-1)+8 (d-2)^2 (n_f-1) z\\
%\\
b_{13} &=& 2 (d-4) (5 d^2-16 d+6)-2 (2 d-7) (d-2) (d-4) z-2 (d-1)
 (3 d-10) (3 d-8) \xi \\
 & & - 2 (d-1) (d-2) (3 d-8)\xi z+2 (3 d-8) (d-2) (d-3) \xi^2 z+4 (3 d-8) (d-3)^2 \xi^2\\
%\\
b_{14} &=& 2160 d-86 d^4-1792 d^2-384+5 d^5+585 d^3+(608 d^2+960-
109 d^3+d^5+2 d^4\\
& & -1304 d)z-2 (d-2) (d-4) (d-6) (2 d-7) z^2-2
(3 d-8) (d-4) (d^3-9 d^2+25 d-18) \xi \\
& & - 2 (3 d-8) (d-3) (d^3-8
d^2+14 d+4) \xi z-2 (d-6) (d-2) (d-4) \xi z^2\\
& & +(3 d-8)(d-3) (d-4) (d^2-9 d+16) \xi^2 z+(3 d-8) (d-3) (d-4) (d^2-9 d+16) \xi^2\\
%\\
b_{15} &=& -16 (n_f-1) (d-2)-16 (n_f-1) (d-2) z\\
%\\
b_{16} &=& -4 (d-4) (3 d^2-11 d+3)+4 (d-4) (d^2-9 d+13) z+8 (d-2)
 (d-4) z^2\\
 & & +2 (d-1) (3 d-10)(3 d-8) \xi+16 (d-1) (d-3)^2 \xi z-
2 (d-1) (d-2) (d-4) \xi z^2 \\
& & -4 (3 d-8) (d-3)^2 \xi^2 z-4 (3 d-8) (d-3)^2 \xi^2\\
%\\
b_{17} &=& -(d-2) (d^2-10 d+28)-(d-2) (d^2-5 d+5) z-(d-2) (d-4)
\xi+(-d+2) \xi z\\
%\\
b_{18} &=& -2 (d-5) (d-3) (d-4)+(-72 d^3-4044 d+988 d^2+4664) z\\
& & + (
-3676 d+3392-140 d^3+1288 d^2) z^2+(-212 d^2+24 d^3+612 d-616)
 z^3 \\
 & &  -2 (d-5)(d-3) (d-4) z^4\\
%\\
b_{19} &=& -15+d+(2 d-2) z-(d-1) (2 d-7) \xi+(-d+1) \xi z+(3-d) \xi^
2
\end{eqnarray*}
\normalsize
\bigskip

{\Large \bf Appendix D}\\

   From the small $q^2$ expansion, as described in appendix B, we get
the following expressions in the limit $z=q^2/m^2 \to 0$
for the renormalized amplitudes $A_{\rm ren}$ and $B_{\rm ren}$.
\small
\begin{eqnarray}
A_0 \z=\z
   C_F \frac{\alpha_s}{4\pi}\Bigl\{
       1 + 3 L_{\mu} + \xi(1+L_{\mu})
       + z(\frac32 + \frac12 \xi )
       + z^2 (\frac12+ \frac16 \xi) \Bigl\}  \nonumber\\
  \z\z
   +(\frac{\alpha_s}{4\pi})^2\Bigl\{
      C_A C_F\Bigl[
            \frac{629}{24}
          + \frac{11}{2} \xi
          + \frac{5}{8} \xi^2
          - \frac{81}{2} S_2
          + L_{\mu}(\frac{313}{12} + \frac72 \xi + \frac34 \xi^2)
          + L_{\mu}^2 (\frac{11}{2}+ \frac34 \xi + \frac14 \xi^2)
                           \nonumber\\
  \z\z
          + \zeta_2(1- \frac32 \xi + \frac12 \xi^2)
       +z\Bigl(
            \frac{403}{24}
          + \frac54 \xi
          + \frac58 \xi^2
          - \frac94 S_2
          + L_{\mu}(\frac{11}{2} + \frac34 \xi + \frac14 \xi^2)
          + \zeta_2(1 + \frac12 \xi)
         \Bigr)
                           \nonumber\\
  \z\z
     +z^2\Bigl(
            \frac{313}{36}
          + \frac23 \xi
          + \frac13 \xi^2
          - \frac14 S_2
          + L_{\mu}(\frac{11}{6} + \frac14 \xi + \frac{1}{12} \xi^2)
          + \frac16 \xi\zeta_2
         \Bigr)\Bigr]
                           \nonumber\\
  \z\z
  + C_F^2\Bigl[
          - 19
          - 10 \xi
          + \xi^2
          + 81 S_2
          + L_{\mu}(- 15 - 2 \xi + \xi^2)
          + L_{\mu}^2(\frac92 + 3 \xi + \frac12 \xi^2)
          + \zeta_2(1 + 6 \xi - \xi^2)
                           \nonumber\\
  \z\z
      + z\Bigl(
          - \frac{57}{4}
          + \xi
          + \frac14 \xi^2
          + \frac92 S_2
          + L_{\mu}(-\frac92 + \frac12 \xi^2)
          + 3 \zeta_2
          \Bigr)
    + z^2\Bigl(
          - \frac{295}{36}
          + \frac12 \xi
          + \frac{1}{12} \xi^2
          + \frac12 S_2
                           \nonumber\\
  \z\z
          + L_{\mu}(-\frac92 - \xi + \frac16 \xi^2)
          + \frac53 \zeta_2
          \Bigr)\Bigr]
  +C_F t\Bigl[
          - 32
          - \frac{20}{3} n_f
         + 162 S_2
         - \frac{20}{3} n_f L_{\mu}
         - 2 n_f L_{\mu}^2
                           \nonumber\\
  \z\z
         + 4 \zeta_2(1-n_f)
      +z\Bigl(
          - 60
          + 252 S_2
          - \frac{16}{3} n_f
          - 2 n_f L_{\mu}
          \Bigr)
                           \nonumber\\
  \z\z
     +z^2\Bigl(
          - \frac{800}{9}
          + 352 S_2
          - \frac{25}{9} n_f
          - \frac23 n_f L_{\mu}
         \Bigr)\Bigr]
    \Bigr\} + O(z^3)
\end{eqnarray}

\begin{eqnarray}
B_0 \z=\z
   C_F \frac{\alpha_s}{4\pi}\xi\Bigl\{
       \frac12+L_{\mu}
       + \frac23 z + \frac14 z^2 \Bigl\}  \nonumber\\
 \z\z
   +(\frac{\alpha_s}{4\pi})^2\Bigl\{
      C_A C_F\Bigl[
            \frac{61}{8}
          + \frac{15}{4} \xi
          + \frac14 \xi^2
          - 27 S_2
          + L_{\mu}(\frac{25}{4} + \frac{11}{4} \xi + \frac12 \xi^2)
          + L_{\mu}^2(\frac34 \xi + \frac14 \xi^2)
                           \nonumber\\
  \z\z
          + \zeta_2(3 - \frac32 \xi + \frac12 \xi^2)
       +z\Bigl(
            \frac{71}{12}
          + \frac{13}{12} \xi
          + \frac12 \xi^2
          - 15 S_2
          + \xi L_{\mu}
          + \frac13 \xi^2 L_{\mu}
          + \zeta_2(\frac{11}{6} + \frac12 \xi)
          \Bigr)
                           \nonumber\\
  \z\z
      +z^2\Bigl(
            \frac{1409}{432}
          + \frac56 \xi
          + \frac{17}{48} \xi^2
          - \frac{133}{12} S_2
          + L_{\mu} (\frac{3}{8} \xi + \frac{1}{8} \xi^2)
          + \zeta_2(1 + \frac16 \xi)
          \Bigr)\Bigr]
                           \nonumber\\
  \z\z
     + C_F^2\Bigl[
            \frac18
          - 8 \xi
          + \frac12 \xi^2
          + 54 S_2
          + L_{\mu}(-\frac32 - 6 \xi + \frac12 \xi^2)
          + \frac12 \xi^2 L_{\mu}^2
          + \zeta_2(-6 + 6 \xi - \xi^2)
                           \nonumber\\
  \z\z
       +z\Bigl(
          - \frac{49}{12}
          - \frac{11}{6} \xi
          - \frac{1}{12} \xi^2
          + 30 S_2
          + L_{\mu}(- 4 \xi + \frac23 \xi^2)
          - \frac43 \zeta_2
          \Bigr)
     +z^2\Bigl(
          - \frac{599}{216}
          - \frac94 \xi
          + \frac{1}{12} \xi^2
          + \frac{133}{6} S_2
                           \nonumber\\
  \z\z
          + L_{\mu}(- 3 \xi + \frac14 \xi^2)
          - \zeta_2
          \Bigr)\Bigr]
     +C_F t\Bigl[
          - 24
          - \frac52 n_f
          + 108 S_2
          - 2 n_f L_{\mu}
       +z\Bigl(
          - \frac{128}{3}
          - \frac43 n_f
          + 168 S_2
          \Bigr)
                           \nonumber\\
  \z\z
     +z^2\Bigl(
          - \frac{1628}{27}
          - \frac12 n_f
          + \frac{700}{3} S_2
          \Bigr)\Bigr]
     \Bigr\} + O(z^3).
\end{eqnarray}
\normalsize
\bigskip

{\Large \bf Appendix E}\\

   Using the asymptotic formulae discussed in appendix B, we obtain
the following expressions in the limit $z=q^2/m^2 \to \infty$
for the renormalized amplitudes $A_{\rm ren}$ and $B_{\rm ren}$
($L_q=\ln(-q^2/\mu^2)$)
\small
\begin{eqnarray}
%  1st line
&&A_{\infty}= \adp C_F \left[ ( 4  - 3L_q + \xi(2- L_q ))
       + \frac{1}{z} (1+L_q+L_{\mu} )(3+\xi)
       - \frac{1}{2z^2}( 3+\xi )
       \right] \nonumber \\
%  2 nd line
&&~~+\left(\adp\right)^2 C_F \left\{
  C_A \left[\frac{1531}{24}-3(7+\xi) \zeta_3 + 10 \xi + \frac{15}{8} \xi^2
  -(\frac{445}{12}+ 5 \xi + \frac54 \xi^2 ) L_q
  \right. \right. \nonumber \\
%  3d line
&&~~+ (\frac{11}{2} + \frac34 \xi + \frac14 \xi^2)L_q^2
  +\frac{1}{z} \left( \frac{349}{12} + \frac52 \xi + \frac34 \xi^2
  + 3(3+\xi)\zeta_3
  + (\frac{181}{12}+ 2 \xi + \frac14 \xi^2)L_q
  \right.  \nonumber \\
%  4th line
&&~~\left.
  - (11+ \frac32 \xi +\frac12 \xi^2   )L_q^2
  + (\frac{313}{12} + \frac72 \xi + \frac34 \xi^2
  - (11+\frac32 \xi +\frac12 \xi^2) L_q
   ) L_{\mu}
               \right)
  \nonumber \\
% 5th line
&&~~+ \frac{1}{z^2} \left( - \frac{14}{3}+ \frac12 \xi
    - \frac14 \xi^2+( 23 + 3 \xi + \frac12 \xi^2)L_q
     +\frac34(9+ \xi)L_q^2
     \right. \nonumber \\
% 6th line
&&~~ \left. \left.
     +( \frac{35}{2}+ \frac94 \xi + \frac14 \xi^2
     +\frac32(9+ \xi)L_q )L_{\mu}
     + \frac34 ( 9 + \xi )L_{\mu}^2
         \right)
         \right] + t \left[ (-\frac{52}{3}+\frac{32}{3} L_q - 2 L_q^2  )n_f
        \right.
        \nonumber \\
&&~~
 + \frac{1}{z} \left(-\frac83 L_q n_f - 24 L_q + 4 L_q^2 n_f
      -\frac{20}{3} n_f
      - (  24 - 4 L_q n_f + \frac{20}{3} n_f )L_{\mu}
              \right) \nonumber \\
&&~~\left.
    +\frac{1}{z^2}\left( 48 - 4 ( n_f+9) L_q +\frac73 n_f
       -2 ( 18+n_f )L_{\mu}
                     \right) \right] \nonumber \\
&&~~+ C_F \left[  13+ 12 \zeta_3 + 8 \xi + \xi^2
       -2(\xi^2+5 \xi  + 6)L_q
       + (\frac92 + 3 \xi+ \frac12 \xi^2  )L_q^2
          \right. \nonumber \\
&&~~
+\frac{1}{z}  \left(  - 16+ 4 \xi + 4 \xi^2 + 12 \zeta_3
    + 6(3+\xi) L_q- ( \xi+3)^2 L_q^2
 \right. \nonumber \\
&&~~\left.
    + ( 27+ 12 \xi + \xi^2  - \xi^2 L_q + 9 L_q )L_{\mu}
    + 6( 3 + \xi )L_{\mu}^2
                  \right)
+\frac{1}{z^2} \left(-\frac{69}{4}-\frac{17}{2} \xi+\frac54 \xi^2 \right.
  \nonumber \\
&&~~\left.  \left. \left.
        - (3- 6\xi - 4\xi^2)L_q - 6 L_q^2
       -(\frac{51}{2}  + 3\xi - \frac72 \xi^2+ 12 L_q )L_{\mu}
       -6 L^2_{\mu}  \right)
             \right] \right\} + o(\frac{1}{z^2}).
\end{eqnarray}

%    g QinfValueB=oneMinusB;
%   QinfValueB =
\begin{eqnarray}
&&B_{\infty}= \adp C_F \xi \left[  1 - L_q + \frac{2}{z}
       + \frac{1}{z^2}  (\frac12 + L_q  + L_{\mu} )\right]
  \nonumber \\
&&~~+ \left(\adp\right)^2 C_F \left\{
 C_A \left[ \frac{41}{4} +\frac{13}{2} \xi + \frac98 \xi^2
 - 3(1+\xi) \zeta_3 - (\frac{25}{4} +\frac72 \xi
 + \frac34 \xi^2)L_q +\frac14(3+\xi)\xi L_q^2  )
    \right. \right.
 \nonumber \\
&&~~
 +\frac{1}{z} \left( \frac{31}{2}  + \frac92 \xi + \xi^2
 + 3(\xi-3)\zeta_3 -\xi(3+\xi) L_q \right)
 \nonumber \\
&&~~ +\frac{1}{z^2} \left(-\frac{27}{2}+ \frac{19}{4} \xi
     + \frac34 \xi^2 + 24 \zeta_3
       - (\frac{17}{4} - \frac{7}{2} \xi - \frac14 \xi^2) L_q
        -(\frac94+\frac34 \xi +\frac12 \xi^2 )L_q^2
   \right.
   \nonumber \\
&&~~\left. \left.
  +(- \frac{17}{4} - \frac12 L_q \xi^2 - \frac92 L_q
       + \frac{17}{4} \xi+ \frac12 \xi^2)L_{\mu}
       +\frac34 ( \xi-3 )L_{\mu}^2 \right) \right]
 \nonumber \\
&&~~+ t \left[  ( 2 L_q - \frac72 )n_f
       +\frac{1}{z}  \left( 18 - 12 L_q - 4 n_f -12L_{\mu} \right)
 \right.
 \nonumber \\
&&~~\left. +\frac{1}{z^2} \left( \frac{332}{9}- n_f  - 2L_q n_f
   +\frac83 L_q + 4 L_q^2 + (\frac83 + 8 L_q - 2 n_f )L_{\mu}
       + 4L_{\mu}^2 \right) \right]
\nonumber \\
&&~~+ C_F \left[-\frac58 - L_q \xi^2 + \frac32 L_q
    +\frac12 \xi^2 L_q^2+\frac{1}{z} \left(-12
     + 10 \xi + 4\xi^2-2\xi(6+\xi) L_q  \right) \right.
\nonumber \\
&&~~
 + \frac{1}{z^2} \left( \frac{11}{4}+ \frac{15}{2} \xi
 + \frac{11}{4}\xi^2 - 24 \zeta_3
 + (19 \xi + 3 \xi^2 - \frac92) L_q -\xi(9+\xi) L_q^2 \right.
\nonumber \\
&&~~\left. \left. \left.
    -(\frac92 - 19 \xi - \frac72 \xi^2 +\xi(6+\xi) L_q)L_{\mu}
    -3 \xi L_{\mu}^2 \right)
       \right]
       \right\}+ o(\frac{1}{z^2}).
\end{eqnarray}

\end{document}